# Effect of Shock Strength on the Radiation of Focusing Shock Wave

Saranyamol V. S.          Mohammed Ibrahim S.

Hypersonic Experimental Aerodynamics Laboratory (HEAL)
Department of Aerospace Engineering, Indian Institute of Technology, Kanpur,
Uttar Pradesh, India, 208016.

**ABSTRACT**

High temperature radiating Air is produced experimentally by focusing a shock wave with the help of a spherically converging test section attached to a shock tube. The converging section concentrates the shock to a point with minimum diffusion losses. A shift in radiation towards the UV region was observed with an increase in the strength of the focusing shock wave. The atomic and molecular emission was observed from the radiation spectrum. Along with the emission from molecules of Air, emissions from contaminations were also observed. The temperature of the radiating gas was estimated using the blackbody radiation curve and was observed to be 13000 K.

## 1  INTRODUCTION

Shock waves are thin discontinuous regions that cause an abrupt rise in fluid properties like pressure, temperature, etc. Converging these high-energy shock waves to a tiny region in space will result in a very high energy concentration. This phenomenon is called shock wave focusing, and it has various applications like inertial confinement fusion [1][2], shock wave lithotripsy, material science [3], ignition techniques [4], etc. As the strength of the shock increases, the focused region will encounter a high enough temperature that the gas in this region will start radiating [5]. Shock focusing resembles the phenomenon of gas bubble sonoluminescence [6], supernova collapse [7], [8] etc., and also finds its place in exploring the study in Richtmyer–Meshkov instability [9].

The shock focusing phenomenon has been of great interest to researchers since 1942. Guderly [10] was the first to do theoretical studies on strong cylindrical and spherical shock waves. He proposed a self-similar solution in an ideal gas flow, which express the radius of a converging shock wave as a function of time. Pioneer experimental study on converging shock waves was done by Perry and Kantrowitz [11], achieving cylindrical shock focusing with a tear-drop insert inside a shock tube. A detailed characterisation and study on cylindrical shock focusing in a shock tube was carried out by Zhai et al. [12], [13]. Several other shock focusing techniques like hemispherical implosion chamber [14], parabolic reflector in shock tube [15], annular shock tube [14], etc., were achieved later.

Spherical shock wave focusing was achieved by Setchell et al. [17] with the help of a uniform cone attached to a shock tube. This procedure of focusing the shock was associated with Mach reflections, resulting in losses to the shock. The losses were overcome by Saillard et al. [18] by passing the shock through a smooth curve. The challenge of focusing a spherical shock with minimum losses was also successfully overcome by Apazidis et al. [19]–[22]. A perfectly contoured converging section helped to smoothly vary the shock profile from planar to spherical with minimum diffusion losses.

When the spherical shock of increased strength passes through the converging section, the test gas is heated and emits radiation. Several researchers have carried out experiments on radiating gas mixtures. Knystautas et al. did studies on acetylene-oxygen gas mixture[23], Saito and Glass [24] did measurements of the hydrogen-oxygen mixture. Spectrometric measurements on a cylindrical converging shock wave in Air were carried out by Matsuo et al. [25]. Shockwaves were generated by detonation inside a circular test chamber. The shock reflected from the wall was focused, and photomultiplier tests were carried out on this radiating region. Comparison with blackbody radiation spectrum was made at wavelengths between 400 and 500 nm, and the temperature were estimated to be between 13000–34000 K

Radiation measurement of a spherical shock focusing has been carried out by Malte [21]. Spherical shock wave focusing is achieved with the help of a converging section attached to a shock tube. They carried out experiments with Argon and Nitrogen as the test gas. The radiation measurement obtained from argon test gas was further analysed, and the temperature of the radiating flow field was obtained by comparing it with the blackbody radiation spectrum. They took spectra at various exposure times and different trigger times, which helped them to obtain spectra at different instants of focusing. A maximum blackbody temperature of up to ~27000K was obtained for the Argon test gas, where the initial shock strength was 3.9.

Experimental studies on radiation measurement and identifying the major radiating species in Air in a shock-focused region are minimal. Numerical studies on shock wave focusing have been reported where the dissociation and recombination of Oxygen and Nitrogen were observed [28][29]. In this backdrop, the present work aims to understand the radiation in Air when it faces a converging spherical shockwave experimentally. The focusing is achieved with the help of a smooth converging section attached to a shock tube. The emission from the radiating gas was studied through emission spectroscopy techniques.

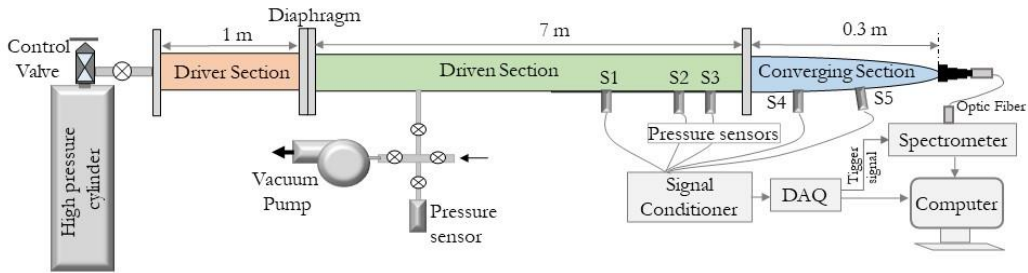

*Figure 1: Schematic diagram of shock focusing setup*

## 2 EXPERIMENTAL METHODOLOGY

### 2.1 Shock Tube Facility

Experiments are performed using the shock tube facility, 'S1 (Vaigai)' at the Hypersonic Experimental Aerodynamics Laboratory (HEAL), Indian Institute of Technology, Kanpur, India. The facility has a 1 m long driver section, a 7 m long driven section, and an 87 mm internal diameter. A converging section of length 0.3 m is attached at the driven section end, which helps to focus the shock wave. The schematic of the facility is shown in Figure 1. The test gas used in the driven section for all the tests was Air, and the gas filled in the driver section was helium. An aluminium diaphragm separates the driver and driven sections.

### 2.2 Shock Focusing facility

The converging section attached to the shock tube transforms the planar shock generated in the shock tube into a smooth spherical shock with minimum loss to the shock front. The internal diameter of the tube is reduced smoothly from 87 mm to 18 mm at the focusing end wall by the converging section. An additional converging cone is inserted here, where the diameter further reduces to 0.6 mm, as shown in Figure 2. The design of the converging section was made according to the geometric relations mentioned in equation 1, which was adapted from Malte [26].

$$x = A \sin \theta$$
$$y = B - R(1 - \cos \theta)$$
(1)

where $\theta$ is ranging from 0 to $0.35\pi$; A= 339.7; B= 43.5, and R= 47.

### 2.3 Instrumentation

Unsteady pressure measurements are carried out using the ICP® pressure sensor of PCB piezotronics, model No-113B22, flush-mounted along the surface of the facility, as can be seen in Figure 1. The pressure data was acquired using NI-USB 6356, a multifunction I/O DAQ device at a rate of 1.0 Mega samples per channel over a duration of 0.25 seconds. The DAQ starts acquiring the signal when the shock reaches sensor S1 (refer to Figure 1). The sensors were connected to the DAQ through a PCB signal conditioner (Model No. 482C05). There are several factors that affect the measurement of the unsteady pressure sensor. There can be an instrumental error in the data acquisition system, error from the connections and sensors, etc. The possibility of human error is also inevitable. Experiments with the same test conditions were repeated three times, and the uncertainty in the measured values was obtained accordingly. This contributes to a ±10% variation in the measured signal [27].

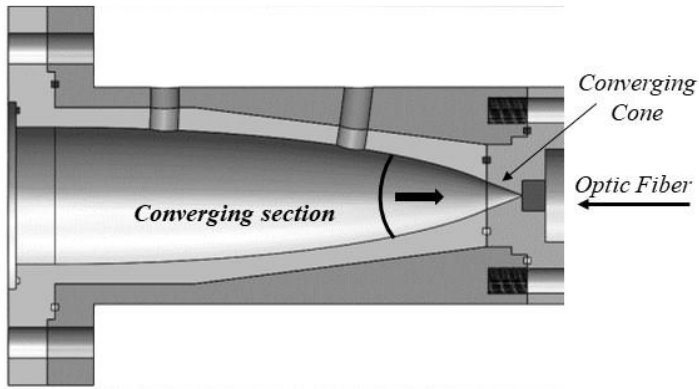

*Figure 2: Schematic of the Overall converging section*

The spectra from the radiating flow field are captured by the spectrometer through an optic fibre cable, as shown in Figure 1. A quartz window is placed in between the focusing point and the optic fibre to avoid damage to the cable. The spectrometer used for current experiments is HO-CT216-3010 from Holmarc®, which captures the spectrum in the UV-VIS-NIR range (340 nm to 1080 nm). The spectrometer acquires data for a minimum exposure time of 1 ms. The spectrometer is triggered with the help of an output voltage signal generated from the DAQ. A trigger delay time was added to the spectrometer if required, depending on the initial shock strength.

The transmittance of the quartz window is estimated before the test run. Figure 4 shows the percentage transmittance of the quartz window. Above 65% transmittance is seen between the wavelength range of 380 nm to 980 nm. The image of the quartz window before and after the impingement of shock is shown in Figure 3. Clearly, the focusing has created a dark patch on the quartz window, which are due to the gas species and other impurities in the test facility deposited on its surface. Black colour deposits on the quartz are due to the carbon contamination generated from the material used in the fabrication of the converging section (Polyoxymethylene). The quartz window was replaced before each experiment.

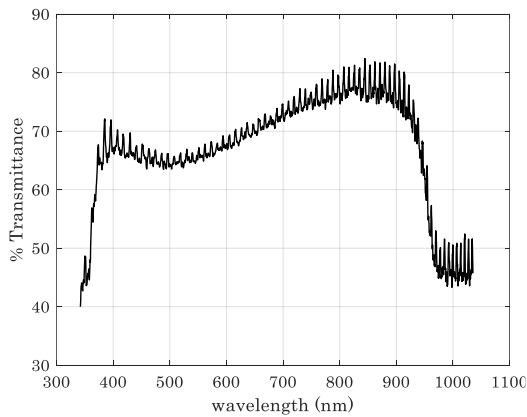

*Figure 4: Transmittance of Quartz*

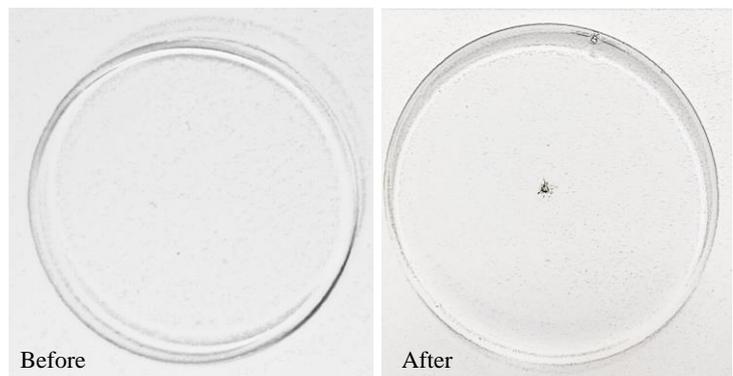

*Figure 3: Quartz window before and after shock impingement*

# 3   EXPERIMENTAL RESULTS

Three sets of experiments were carried out in order to study the effect of shock strength on the radiating gas. The test gas used was Air and the different shock strengths achieved was $M_s$= 2.2, 3.3, and 4.1. The details of test conditions are mentioned in table 1.

*Table 1: Test conditions*

| Shock Mach number ($M_S$) | Driven gas Pressure | Nomenclature |
|---|---|---|
| 2.2 | 0.025 MPa | $M_S$ 2.0 |
| 3.3 | 0.025 MPa | $M_S$ 3.0 |
| 4.1 | 0.01 MPa | $M_S$ 4.0 |

## 3.1   Pressure sensor signals

Unsteady pressure is measured at five locations on the facility, three are flush mounted on the shock tube, and two are on the converging section. The pressure at each location for $M_S$ 4.0 is shown in

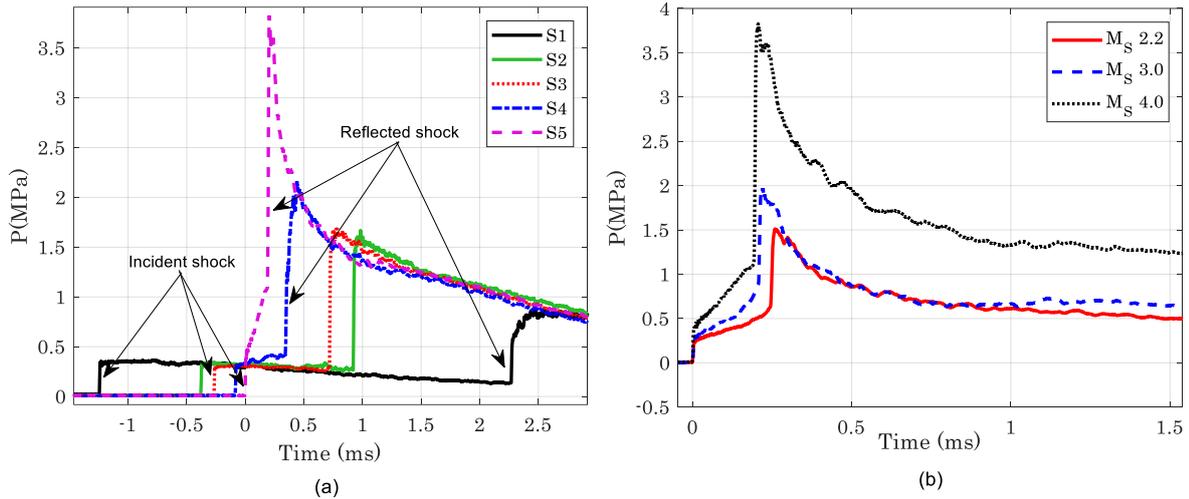

*Figure 5: Pressure measurement results are shown here. (a) shows the pressure measured at all the sensor locations for $M_S$ 4.0. (b) shows the pressure measured at location S5 for all the three shock Mach number cases.*

Figure 5 (a). The incident shock, as well as the reflected shock, is marked in the figure. The gradual rise in pressure is observed as the shock moves forward. The strength of the shock increases as it moves through the converging section, resulting in a rise in pressure across it. Maximum pressure of 3.83 MPa is obtained with the reflected shock at the location S5 for $M_S$ 4.0. Behind the reflected shock, a drop in pressure is observed owing to the rapid expansion of the high-temperature, high-pressure gas generated after the shock focusing event. The shock Mach number reduces as the reflected shock passes through a diverging section. The pressure measured at location S5 for all three shock cases is depicted in Figure 5 (b). The magnitude of pressure value is observed to have increased with shock Mach number.

## 3.2   Spectroscopic results

The radiation measurement of the test gas at the focusing point is obtained through a spectrometer, as discussed earlier. The spectrum is obtained for three different shock strengths to study its effect on radiation in Air in the focused region, and the result is shown in Figure 6. The spectra are normalised with respect to the maximum value of itself. The obtained spectrum can be classified into three regions: UV (350-380 nm), Visible (380-750 nm), and NIR (starting from 750 nm). It was observed that the radiation intensity shifted towards the UV region from NIR as the shock strength was increased. In the NIR region, the radiation intensity was the same for all three shock strengths except for the peak at 767 nm, for which the weakest shock, $M_S$ 2.0, has the strongest intensity. Also, in the red end of the visible spectrum, the radiation intensities were all the same for the three test cases. In the visible region, starting from 540 nm to 650 nm, the radiation intensities were comparable for $M_S$ 2.0 and $M_S$ 3.0 test cases. However, the intensity was lower for the strongest case; $M_S$ 4.0. Below 540 nm in the visible region and in the UV region, radiation intensity increased with increasing shock strength and was highest for the $M_S$ 4.0 test case. The shift in radiation intensity from NIR to UV with increasing shock strength is attributed to the fact that stronger shock waves result in higher temperature in the focusing region [28][29], thereby enabling electronic excitation and subsequent relaxation to ground states of a large number of atomic and molecular species.

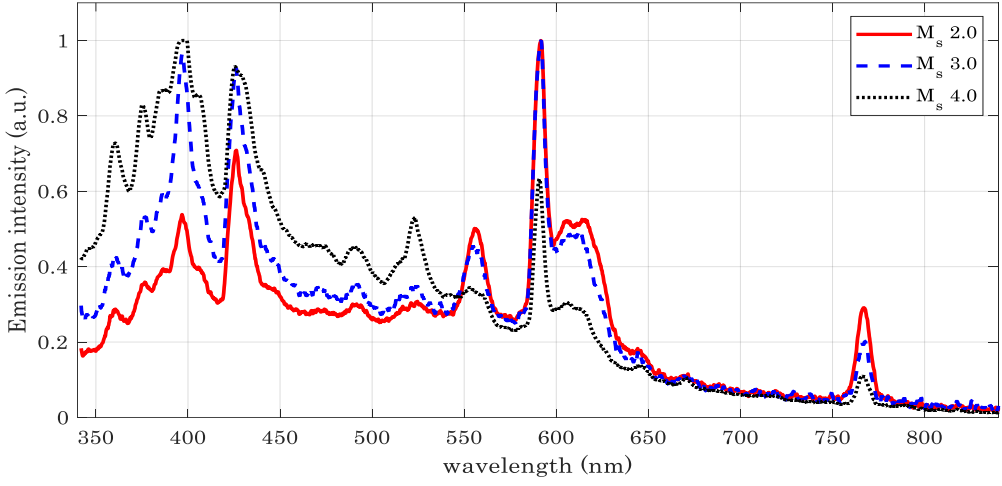

*Figure 6: Radiative spectrum showing the effect of shock Mach number.*

The atomic and molecular species emitting radiation are identified and are shown in Figure 7. The emission from molecules $N_2$, CN radical, and molecular ions $N_2^+$ are observed in the range of 300 nm to 500 nm [30][31][32]. $N_2^+$ ($B^2\Sigma_u^+ - X^2\Sigma_g^+$) (First negative system) emits at 356.4 nm and 391.4 nm, respectively, and $N_2$ ($C^3\Pi_u - B^3\Pi_g$) (second positive system) emits in the range of 340-382 nm [32] [33]. Three different CN violet ($B^2\Sigma^+ - X^2\Sigma^+$) bands ($\Delta v$=0, +1, and -1) radiates in the range of 330 nm to 425 nm [30][33]. Molecular CN emissions are due to the Carbon from polyoxymethylene, the material used in the converging section, combined with Nitrogen [34]. Atomic Oxygen emission lines (O I and O II) are observed at 604 nm, 615 nm, and 645 nm [36]. Atomic Nitrogen emission (N II) is observed at 648 nm [30]. In the range of 700 nm to 750 nm, several small emissions are observed which are due to Atomic Nitrogen and Atomic Oxygen [34].

The electronic transition of Na I causes radiation at 589 nm and 589.5 nm [35]. It is possible that both the radiation lines got merged and are observed as a single peak here due to the low resolution of wavelength and pixel of the spectrometer. Other atomic emissions like Lithium (Li I) at 427.3 nm and 671.3 nm [37], Potassium (K I) at 766.8 nm [38][37], Aluminium at 400 nm (Al III), 671.3 nm (Al I and Al II) [38], and 766 nm (Al III) [35] are also observed. AlO band emission ($B^2\Sigma^+ - X^2\Sigma^+$) is observed at a range of 450 nm to 540 nm [39][21]. Molecular emissions bands from CaO transitions ($d^3\Delta_2 - a^3\Pi_1$), ($e^3\Sigma - a^3\Pi_1$), ($c^3\Sigma^+ - a^3\Pi_1$), ($C^1\Sigma^+ - a^3\Pi_1$), ($D^1\Delta - a^1\Pi_1$), ($c^3\Sigma_1 - a^3\Pi_0$), ($c^3\Sigma^+ - a^3\Pi_2$) etc are also observed in the obtained spectrum as seen in the figure [39]–[41]. Lithium and Potassium emissions are common contaminants in a shock tube, which are present due to the grease used at several locations in the facility. These might have evaporated and radiated due to high temperatures. Aluminium is also a strong contamination since the diaphragm used here is made of Aluminium. In addition to these, Sodium (Na), Iron, Chromium, etc., are also impurities found in a shock tube [42] [43].

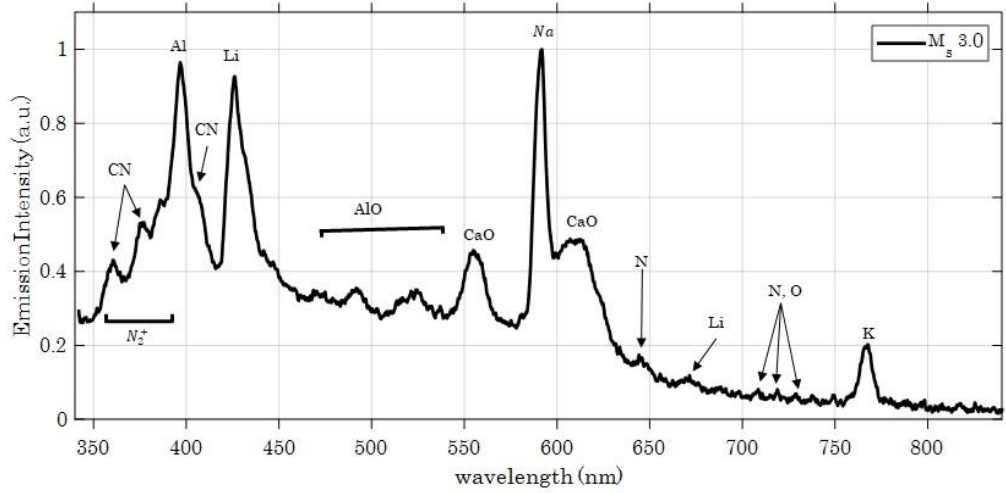

Figure 7: Atomic and molecular emission in radiating air

### 3.3 Comparison with Blackbody radiation

It is observed from the spectra that the emission from the gas species $N_2$ and $O_2$; the primary gas molecules present in Air are overlapped by the emission from the contaminants present in the test facility. Nevertheless, the emission from $N_2^+$, CN, CaO and AlO clearly indicate that the primary gas molecules have undergone dissociation and ionisation, suggesting the temperature in the focusing region is significantly higher. In order to get an estimate of the temperature, the spectra are compared with the blackbody radiation function [44][19]. This is obtained by fitting the radiation spectrum with Planck's Law shown in equation 2.

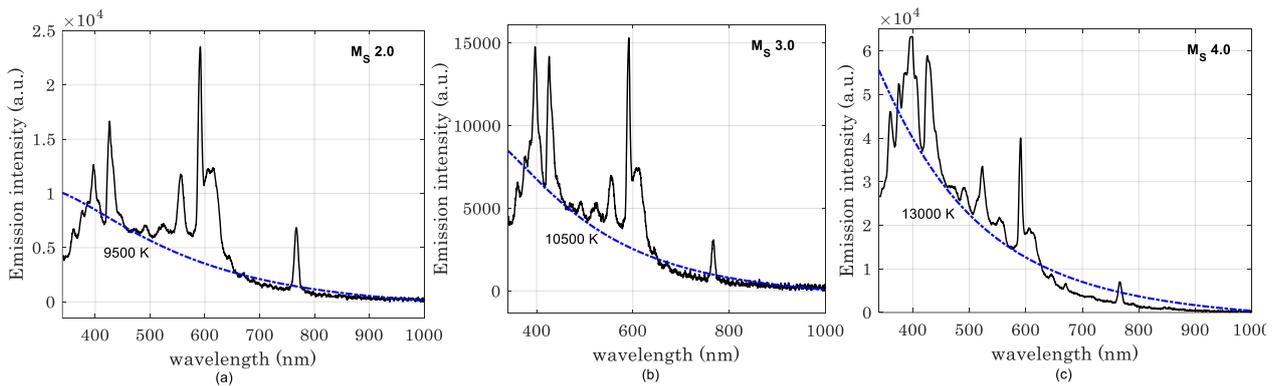

Figure 8: Blackbody fit with the measured spectrum. The experimentally obtained spectra is shown in black solid line and the fitted blackbody curve is shown in blue dotted line.

Here $I(\lambda, T)$ is the intensity as a function of wavelength ($\lambda$) and Temperature (T), 'h' is Planck's constant, 'c' is the speed of light in the medium, 'k' is Boltzmann's constant and 'A' is the fitting parameter [19].

$$I(\lambda, T) = \frac{Ahc^2}{\lambda^5} \frac{1}{exp(hc/\lambda kT) - 1} \qquad 2$$

The baseline of the measured spectrum follows a curved path, which is high in the UV region and reduces gradually towards the NIR region. The blackbody temperature curve is fitted along the baseline of the obtained spectrum, and the result is shown in Figure 8. The temperature obtained from the blackbody curve for the test case $M_S$ 2.0 is 9500±100 K, for $M_S$ 3.0 is 10500±100 K and for $M_S$ 4.0 is 13000 ±100K. The temperature generated in the focusing region is observed to be increasing with an increase in shock strength.

## 4   CONCLUSION

Experiments were carried out with three different initial shock strengths to study the effect of shock strength and radiative emission of the shock-focused gas. It was observed that with the increase in shock strength, the radiation intensity shifted to the UV region. Apart from the primary gas molecular emission, which is Nitrogen and Oxygen, we also observed emissions from contaminants. It is the contaminants which are overlapping with the gas molecular/atomic emissions. Emissions from Carbon, Sodium, Aluminium, Lithium, etc., were observed, which are inevitable in test facilities like a shock tube. Nevertheless, the temperature in the shock-focused region will reach a maximum of 13,000K. This temperature causes dissociation and recombination of Oxygen and Nitrogen, resulting in the formation of Oxides of Aluminium, Calcium, and CN formation.

### ACKNOWLEDGMENTS

The authors would like to thank the Science and Engineering Research Board (SERB), India, for supporting this research work under the Early Career Research Award, ECRA/2018/000678.